\documentclass{amsart}
\usepackage{graphicx}
\usepackage{epsf}
\def\be{\begin{equation}}
\def\ee{\end{equation}}
\def\ba{\begin{array}}
\def\ea{\end{array}}
\def\bea{\begin{eqnarray}}
\def\eea{\end{eqnarray}}
\def\drm{{\mathrm d}}
\def\erm{{\mathrm e}}
\def\dps{\displaystyle}
\def\rt{\rightarrow}
\def\ul{\underline}
\def\n{\nu}
\def\t{\tau}
\begin{document}

\vspace{-4truecm} %
{}\hfill{DSF$-$36/2005} %
%physics/0512174 %
\vspace{1truecm}

\title{Una lezione particolare di Ettore Majorana}%
\author{S. Esposito}%
\address{{\it S. Esposito}: Dipartimento di Scienze Fisiche,
Universit\`a di Napoli ``Federico II'' \& I.N.F.N. Sezione di
Napoli, Complesso Universitario di M. S. Angelo, Via Cinthia,
80126 Napoli ({\rm Salvatore.Esposito@na.infn.it})}%

%\thanks{}%
%\subjclass{}%
%\keywords{}%

%\date{}%
%\dedicatory{}%
%\commby{}%
%----------------------------------------------------------------
\begin{abstract}
Si riporta per la prima volta il testo critico completo di un
manoscritto di Ettore Majorana, conservato insieme agli appunti
per le lezioni tenute all'Universit\`a di Napoli e redatto
probabilmente per un seminario svolto in quella Universit\`a.
Alcuni passaggi del testo rivelano, in forma latente, una
interpretazione fisica della meccanica quantistica che anticipa di
molti anni l'approccio di Feynman in termini di integrali sui
cammini, indipendentemente dalla formulazione matematica
sottostante ad esso.
\end{abstract}

\maketitle

%----------------------------------------------------------------
\section{Introduzione}

\noindent L'interesse per le lezioni del corso di Fisica teorica,
svolte da Ettore Majorana all'Universit\`a di Napoli nel 1938, \`e
stato recentemente ravvivato dal ritrovamento del Documento Moreno
\cite{moreno}, una trascrizione fedele da parte di Eugenio Moreno
degli appunti delle lezioni preparati dallo stesso Majorana,
comprensiva di alcune lezioni precedentemente ignote e i cui
originali sembrano essere andati dispersi. Gli appunti originali
sono conservati a Pisa, e furono riprodotti anastaticamente alcuni
anni or sono \cite{Bibliopolis}, includendovi anche alcuni fogli
che si \`e voluto interpretare \cite{Senatore} \cite{Cabibbo} come
gli appunti preliminari per la lezione successiva alla
ventunesima, che Majorana avrebbe svolto se non fosse scomparso.
Sebbene il carattere ``preliminare'', ossia di appunti di uso
personale da non consegnare direttamente agli studenti, sia
ampiamente attestato dalle inusuali numerose cancellature che vi
sono presenti\footnote{La presenza di una lista degli argomenti da
trattare \`e, invece, tipico degli scritti (personali) di
Majorana. Si veda ad esempio i suoi ``Volumetti'', pubblicati in
\cite{volumetti}.}, un'analisi attenta dei contenuti ivi svolti
non sembra indicare che tali appunti siano stati redatti per il
corso di Fisica teorica n\`e, tantomeno, preparati per la lezione
imediatamente successiva alla ventunesima. Infatti, come si pu\`o
constatare facilmente, tali appunti svolgono argomenti avanzati e
niente affatto collegati a quanto trattato da Majorana nelle sue
ultime tre lezioni, in cui egli aveva appena introdotto le basi
della meccanica ondulatoria e della sua interpretazione
statistica. Qui si presuppongono, invece, conoscenze che il
docente non aveva certamente ancora fornito nel suo corso e si
sviluppano delle applicazioni particolari che non possono
riguardarsi come attuazioni immediate del nuovo formalismo da poco
avviato (come la teoria del legame molecolare). \\
D'altra parte, lo stile adottato da Majorana \`e completamente
diverso da quello impiegato nelle altre lezioni rivolte agli
studenti, in cui l'espressione matematica dei concetti fisici ha
un ruolo rilevante, e la sua forma discorsiva si avvicina
generalmente a quella adoperata nella prolusione al corso (rivolta
principalmente ai docenti dell'Universit\`a di Napoli
\cite{moreno}). In particolare sebbene negli appunti considerati
non siano stati svolti tutti gli argomenti (ovvero: le
applicazioni) riportati nella lista iniziale, la frase conclusiva
di Majorana \`e alquanto perentoria e veramente tipica di una
conferenza generale: ``La meccanica quantistica apre la strada per
l'unificazione logica di tutte le scienze che hanno per oggetto
comune il mondo inorganico''. Sebbene non vi siano attualmente
prove documentate dirette, e in ogni caso la questione presenti
poco interesse per quanto si vuole discutere in questo luogo, per
concretezza riterremo qui che tali appunti manoscritti siano stati
preparati per un seminario o una conferenza generale. Se tale
ipotesi trovasse, poi, ulteriore credito, ci si potrebbe anche
spingere a presumere che tale conferenza sia stat verosimilmente
sollecitata dal direttore dell'Istituto di Fisica
dell'Universit\`a di Napoli, Antonio Carrelli, che in quel periodo
lavorava insieme con i suoi collaboratori a questioni sperimentali
di Fisica molecolare \cite{carrelli}, e che quindi il nucleo
centrale della conferenza fosse, come effettivamente appare,
l'interpretazione teorica del legame molecolare nella meccanica
quantistica. Una ipotetica data per lo svolgimento di tale
conferenza potrebbe, allora, essere collocata nella seconda met\`a
di gennaio del 1938, ossia poco dopo l'arrivo di Majorana a
Napoli, e comunque non lontano dal 3 febbraio, quando egli svolse
la lezione N.9, dove affront\`o il problema dello spettro degli
atomi con due elettroni di valenza (da un punto di vista
fenomenologico) e in cui si pu\`o riscontrare una discussione un
po' pi\`u dettagliata delle forze di risonanza (in ambito atomico
piuttosto che molecolare; ma nel manoscritto in
esame si rimanda anche a quello). \\
L'attenzione del lettore, tuttavia, non vuole essere qui rivolto
all'aspetto {\it storico}, che pure pu\`o avere una notevole
importanza, trattandosi di un lavoro scritto comunque poco tempo
prima della scomparsa dell'autore, bens\`i al contenuto {\it
scientifico} del testo. Sebbene, come accennato sopra, l'argomento
principale della dissertazione debba considerarsi l'applicazione
della meccanica quantistica alla teoria del legame molecolare,
l'interesse scientifico di essa \`e maggiormente incentrato sulla
interpretazione che Majorana offre di alcuni fondamenti della
(nuova) Teoria dei Quanti (il concetto di stato quantico) e
dell'applicazione diretta di essa a un caso particolare (appunto
il legame molecolare). Una lettura accorta del manoscritto,
infatti, non solo tradisce una particolare profondit\`a di ingegno
dell'autore nel trattare un argomento cardine della nuova
meccanica ma, tenendo presente che esso fu scritto nel 1938,
rivela anche un notevole anticipo (di almeno una decina di anni)
nell'uso che ne viene fatto, cosa d'altronde quasi usuale nel caso
di Majorana (basta citare, come solo esempio, il caso dell'atomo
di Thomas-Fermi \cite{fermithomas}; si veda, comunque, anche \cite{volumetti}). \\
Tale circostanze fu gi\`a intuita alcuni anni or sono da N.
Cabibbo \cite{Cabibbo}, il quale vide nella presunta futura
lezione di Majorana una vaga e approssimata anticipazione
dell'idea sottostante alla interpretazione di Feynman della
meccanica quantistica in termini degli integrali sui cammini ({\it
path integral approach}). Uno studio pi\`u analitico, condotta
sull'edizione critica dello scritto di Majorana (non disponibile a
quel tempo, ma riportato alla fine del presente articolo per la
prima volta), rivela invece delle interessanti sorprese, sulle
quali si fisser\`a qui l'attenzione. Sebbene la particolare
semplicit\`a e chiarezza espressiva del testo (molto comune nelle
opere di Majorana \cite{volumetti}) possa far apparire come
superflui ulteriori commenti, per un suo accurato inquadramento
verr\`a nel prossimo paragrafo premessa una breve discussione
delle idee fondamentali della teoria degli integrali sui cammini,
a cui seguir\`a una semplice esposizione in cui solo si
evidenzieranno i passaggi cruciali a tale riguardo nello scritto
di Majorana.

\section{La formulazione della meccanica quantistica
con gli integrali sui cammini}

\noindent Il postulato generale su cui si basa la meccanica
quantistica stabilisce, come noto, che lo ``stato'' di un certo
sistema fisico possa rappresentarsi mediante una grandezza
complessa $\psi$, considerata come vettore (normalizzato) di un
opportuno spazio di Hilbert associato al sistema fisico, che
contiene tutte le possibili informazioni sul sistema
\cite{caldirola}. L'evoluzione temporale del vettore di stato \`e
governata dall'equazione di
Schr\"odinger, che pu\`o scriversi sotto la forma generale %
\be \label{e1} %
i \hbar \frac{\drm \psi}{\drm t} = H \, \psi ,
\ee %
dove $H$ \`e l'operatore hamiltoniano del sistema considerato. Lo
stato iniziale al tempo $t_0$ \`e specificato dalla scelta di
questo tra i possibili autostati di un insieme completo di
operatori che commutano tra di loro e con l'hamiltoniana, mentre
la stessa hamiltoniana $H$, tramite l'equazione (\ref{e1}),
determina lo stato del sistema a un istante successivo $t$.
L'evoluzione dinamica del sistema \`e allora completamente
determinata allorquando si calcoli l'ampiezza di transizione dallo
stato $S_0$ al tempo $t_0$ allo stato $S$ al tempo $t$. \\
Come ci si pu\`o facilmente convincere, dunque, la descrizione
usuale della meccanica quantistica di un sistema fisico \`e
fortemente incentrata sul ruolo dell'hamiltoniana $H$ e, come
conseguenza, la variabile temporale gioca un ruolo centrale in
tale descrizione. Una siffatta dissimmetria tra le variabili
spaziali e temporale \`e, ovviamente, insoddisfacente alla luce
dei postulati della Teoria della Relativit\`a, e di ci\`o se ne
rese conto per primo Dirac nel 1932 \cite{dirac}, il quale
avanz\`o l'idea di riformulare l'intera meccanica quantistica in
termini lagrangiani piuttosto che hamiltoniani\footnote{Si
osservi, tuttavia, che la teoria quantistica dei campi d'onda era
gi\`a stata formulata mediante l'introduzione di una funzione
lagrangiana a cui applicare un opportuno principio variazionale.
Si veda, come punto di riferimento, il libro classico di
Heisenberg \cite{Heisenberg}, a cui spesso faceva riferimento nei
suoi studi anche Majorana, come ad esempio per lo sviluppo della
teoria relativistica delle particelle con spin arbitrario
\cite{teorel} o della teoria simmetrica dell'elettrone e del
positrone \cite{elpos}.}. \\
Il punto di partenza nella riflessione di Dirac \`e quello di
sfruttare un'analogia, a livello quantistico, con la funzione
principale di Hamilton nella meccanica classica \cite{goldstein}.
In base a ci\`o, l'ampiezza di transizione per passare da uno
stato individuato dalla configurazione spaziale $q_a$ al tempo
$t_a$ ad uno stato individuato dalla configurazione spaziale $q_b$
al tempo $t_b$ \`e scritta come:%
\be \label{e2}%
\langle q_b | q_a \rangle \sim \erm^{\dps \frac{i}{\hbar} S} =
\erm^{\dps \frac{i}{\hbar} \int_{t_a}^{t_b} \!\! L \drm t}
\ee%
dove $L$ \`e la lagrangiana del sistema e $S[q]$ il funzionale
d'azione. La relazione precedente, tuttavia, non pu\`o essere
considerata un'uguaglianza, almeno fintantoch\'e l'intervallo
temporale fra $t_a$ e $t_b$ resta finito, in quanto condurrebbe a
risultati non corretti (e di ci\`o era ben conscio lo stesso
Dirac, che nel suo articolo produsse tutta una serie di
assunzioni, senza per\`o motivate giustificazioni per esse).
Infatti, supponendo di dividere la regione d'integrazione in
(\ref{e2}) in $N$ intervalli, $t_a = t_0 < t_1 < t_2 < \dots <
t_{N-1} < t_N = t_b$, l'ampiezza di transizione si potrebbe
scrivere come un prodotto di termini, %
\be \label{e3} %
\langle q_b | q_a \rangle = \langle q_b | q_{N-1} \rangle \langle
q_{N-1} | q_{N-2} \rangle \cdots \langle q_2 | q_1 \rangle \langle
q_1 | q_a \rangle ,
\ee %
mentre \`e ben noto, usando le relazioni di completezza, che la
formula corretta contiene le integrazioni sulle regioni
intermedie: %
\bea %
\langle q_b | q_a \rangle &=& \langle q_b | \! \int \!\! \drm
q_{N-1} \, | q_{N-1} \rangle \langle q_{N-1} | \dots \! \int \!\!
\drm q_{1} \, | q_{1} \rangle \langle q_{1} | q_a \rangle \nonumber \\
&=& \! \int \!\! \drm q_1 \drm q_2 \cdots \drm q_{N-1} \, \langle
q_b | q_{N-1} \rangle \cdots \langle q_2 | q_1 \rangle \langle q_1
| q_a \rangle . \label{e4}
\eea %
L'intuizione di Feynman \cite{feynman}, palesata circa dieci anni
dopo l'articolo di Dirac, fu allora quella di ritenere che la
(\ref{e2}) valesse come uguaglianza, a meno di una costante
moltiplicativa $A$, solo per transizioni tra stati separati da un
intervallo di tempo {\it infinitesimo}. In tal caso, dunque,
utilizzando la formula corretta (\ref{e4}), si ottiene la ben nota
espressione di Feynman per l'ampiezza di transizione fra due stato
generici: %
\be \label{e5} %
\langle q_b | q_a \rangle = \!\!\!\!\!\!\!\!\!\!\!\!\!\!\!\!\!\!
\lim_{\scriptsize ~~~~~~~~~~~ \ba{l} N \rt \infty \\
t_b - t_a \, {\rm finito} \ea} \!\!\!\! A^N \! \int \!\! \drm q_1
\drm q_2 \cdots \drm q_{N-1} \, \erm^{i S/\hbar} \; \equiv \; \int
\! {\rm D}q \, \erm^{i S/\hbar} .
\ee %
Il significato della formula precedente pu\`o comprendersi nel
modo seguente. Nelle integrazioni presenti, che si \`e indicato
genericamente con $\int \! {\rm D}q$, gli estremi $t_a$ e $t_b$
dell'intervallo di integrazione sono tenuti fissi, mentre si
integra sull'intero spazio rispetto rispetto ai punti intermedi.
Poich\`e ogni configurazione spaziale $q_i$ dei punti intermedi
corrisponde ad una data traiettoria dinamica che unisce il punto
iniziale $t_a$ con quello finale $t_b$, l'integrazione su tutte
queste configurazioni equivale allora a sommare su {\it tutti} i
possibili cammini che collegano il punto iniziale con quello
finale. In altre parole la formula di Feynman con gli integrali
sui cammini evidenzia che l'ampiezza di transizione tra uno stato
iniziale e uno stato finale pu\`o essere espresso come una somma
su tutti i cammini, con estremi fissati, del fattore
$\erm^{iS[q]/\hbar}$. Tale risultato, da un lato, non \`e affatto
sorprendente se si pensa al fatto che, nella meccanica
quantistica, se un dato processo pu\`o avvenire secondo modalit\`a
diverse, l'ampiezza di transizione \`e data dalla somma delle
ampiezze individuali corrispondenti a {\it tutti} i possibili modi
in cui il processo pu\`o avvenire. Ci\`o risulta molto chiaro, ad
esempio, dal classico esperimento con un fascio di elettroni che
attraversa uno schermo con una doppia fenditura. La struttura
interferenziale che si osserva dietro lo schermo ad una certa
distanza da esso, che sembrerebbe indicare che uno stesso
elettrone sia passato attraverso entrambe le fenditure, \`e
infatti spiegabile ammettendo che la probabilit\`a che l'elettrone
passi dalla sorgente allo schermo di raccolta  attraverso la
doppia fenditura sia ottenuto sommando su {\it tutti} i possibili
percorsi dell'elettrone. L'approccio on gli integrali sui cammini,
dunque, contiene al suo interno i principi fondamentali della
meccanica quantistica. Tuttavia ci\`o che in esso \`e cruciale e
inaspettato \`e che la somma \`e eseguita sul fattore di fase
$\erm^{iS[q]/\hbar}$, generato dall'azione classica $S[q]$. \\
Uno dei migliori pregi dell'approccio di Feynman alla meccanica
quantistica \`e la possibilit\`a di ottenere in modo estremamente
chiaro il passaggio alla meccanica classica nel limite $\hbar \rt
0$ (gli altri pregi essendo la sua versatile applicabilit\`a alla
quantizzazione dei campi nelle teorie di gauge abeliane, non
abeliane e con rottura spontanea della simmetria di gauge).
Infatti, per grandi valori di $S$ rispetto ad $\hbar$, il fattore
di fase nella (\ref{e5}) subisce delle grandi fluttuazioni e
contribuisce quindi con termini che si mediano a zero. Da un punto
di vista matematico, dunque, risulta chiaro che nel limite $\hbar
\rt 0$ il contributo dominante alla (\ref{e5}) viene fuori nel
caso in cui il fattore di fase non cambia molto o, ci\`o che \`e
lo stesso, quando l'azione $S$ \`e stazionaria. Tale risultato \`e
appunto ci\`o che si riscontra nel caso della meccanica classica,
dove le traiettorie dinamiche classiche si ottengono dal principio
di minima azione. Questa circostanza fu notata per primo da Dirac
\cite{dirac}, il quale si rese ben conto del ruolo molto
importante giocato dal funzionale d'azione. \\
L'interpretazione intuitiva del passaggio al limite classico \`e
molto semplice. Nello spazio $q,t$ si consideri un dato cammino
che sia lontano dalla traiettoria classica $q_{\rm{cl}}(t)$;
poich\`e $\hbar$ \`e piccolo l fase $S/\hbar$ lungo tale cammino
sar\`a abbastanza grande. Per ognuno di tali cammini ce ne sar\`a
uno vicino, infinitamente prossimo ad esso, dove l'azione $S$
cambier\`a solo per una piccola quantit\`a, ma poich\`e questa \`e
moltiplicata per una costante ($1/\hbar$) molto grande, la fase
risultante avr\`a un valore altrettanto grande. In media tali
cammini daranno un contributo nullo nella somma (\ref{e5}). Vicino
la traiettoria classica $q_{\rm{cl}}(t)$, invece, poich\`e
l'azione \`e stazionaria, passando ad un cammino infinitamente
prossimo a quello classico l'azione non cambier\`a affatto, per
cui i i corrispondenti contributi alla (\ref{e5}) si sommeranno
coerentemente, e conseguentemente si otterr\`a l'apporto dominante
per $\hbar \rt 0$. In tale approccio, dunque, nel limite $\hbar
\rt 0$ viene selezionata la traiettoria classica non perch\`e essa
contribuisca maggiormente all'evoluzione dinamica del sistema, ma
piuttosto perch\`e vi sono cammini infinitamente prossimi ad essa
che danno contributi che si sommano coerentemente. La regione di
integrazione \`e allora effettivamente molto sottile nel caso di
sistemi classici, mentre diviene molto larga per sistemi
quantistici. Di conseguenza il concetto stesso di orbita, che nel
caso classico \`e ben definito, in situazioni quantistiche viene a
perdere il suo significato, come ad esempio nel caso di un
elettrone che ruota intorno al nucleo di un atomo.

\section{Il contributo di Majorana}

\noindent La discussione presentata sopra dell'approccio alla
meccanica quantistica in termini di integrali sui cammini si \`e
incentrata volutamente sull'aspetto matematico piuttosto che su
quello fisico, in quanto \`e stato esattamente questo il processo
storico che ha portato dall'idea originaria di Dirac del 1932 alla
formulazione di Feynman negli anni '40 (basta per questo
consultare gli articoli originali in \cite{dirac} e
\cite{feynman}). D'altra parte \`e stato proprio lo sviluppo del
formalismo matematico che ha permesso, successivamente, la
notevole sua interpretazione fisica accennata nel paragrafo
precedente. \\
Ritornando allo scritto di Majorana qui in esame, si scorge subito
che {\it nulla} dell'aspetto matematico del peculiare approccio
alla meccanica quantistica \`e presente in esso. Tuttavia una
lettura attenta rivela in maniera altrettanto evidente la presenza
in esso dei fondamenti {\it fisici} della
stessa. \\
Il punto di partenza di Majorana \`e il voler ricercare una
formulazione quanto pi\`u significativa e chiara del concetto di
stato quantico. E, naturalmente, nel 1938 la polemica con le
concezioni della vecchia teoria dei quanti \`e ancora aperta. \\
\begin{quote}
 Secondo la teoria di Heisenberg uno stato quantico
corrisponde non a una soluzione stranamente privilegiata delle
equazioni classiche ma piuttosto a un complesso di soluzioni
differenti per le condizioni iniziali e anche per l'energia, ci\`o
che si presenta come energia esattamente definita, dello stato
quantico corrispondendo a una specie di media presa sulle infinite
orbite classiche raccolte in esso. Gli stati quantici si
presentano cos\`i come i minimi complessi statistici di movimenti
classici \ul{poco differenti} accessibili all'osservazione. Questi
complessi statistici minimi non si lasciano ulteriormente dividere
a causa del {principio di indeterminazione}, enunciato dallo
stesso Heisenberg, che vieta la misura esatta contemporanea della
posizione e della velocit\`a di un particella e cos\`i la
fissazione della sua orbita.
\end{quote}
Si osservi che le ``soluzioni differenti per le condizioni
iniziali'' corrispondono, nel linguaggio di Feynman del 1948,
proprio ai diversi cammini di integrazione; infatti le diverse
condizioni iniziali sono comunque sempre riferite ad uno stesso
tempo iniziale ($t_a$), mentre il determinato stato quantico in
oggetto corrisponde ad un fissato tempo finale ($t_b$). La
questione introdotta dei ``movimenti classici \ul{poco
differenti}'' (la sottolineatura \`e di Majorana), secondo la
precisazione, immediatamente seguente, che ne fornisce il
principio di indeterminazione di Heisenberg, \`e pertanto
palesemente connessa a quella dell'estensione sufficientemente
ampia della regione d'integrazione in (\ref{e5}) per sistemi
quantistici, che viene cos\`i ad essere intimamente legata ad un
principio fisico fondamentale. \\
Il punto cruciale della formulazione di Feynman della meccanica
quantistica \`e, come si \`e visto, il considerare {\it tutti} i
possibili cammini che uniscono il punto iniziale con quello
finale, e non solo quelli corrispondenti alle traiettorie
classiche. Nel testo di Majorana, dopo aver discusso un
interessante esempio relativo all'oscillatore armonico, l'autore
rimarca:
\begin{quote}
 Naturalmente la corrispondenza fra stati quantici e complessi
di soluzioni classiche non \`e che approssimata poich\`e le
equazioni della dinamica quantistica sono in generale indipendenti
dalle corrispondenti equazioni classiche e indicano una reale
modificazione delle leggi meccaniche oltre che un limite posto
alle possibilit\`a di osservazione; ma essa \`e meglio fondata
della rappresentazione degli stati quantici mediante orbite
quantizzate e pu\`o essere utilmente impiegata in considerazioni
di carattere qualitativo.
\end{quote}
E, ancora pi\`u esplicitamente, in un luogo pi\`u avanti, si dice
che la funzione d'onda ``\`e associata nella meccanica quantistica
ad ogni stato possibile dell'elettrone''. Tale riferimento, che
superficialmente potrebbe essere interpretato nel senso usuale che
la funzione d'onda contiene tutte le informazioni del sistema
fisico, \`e invece da considerarsi secondo l'accezione di Feynman,
alla luce della approfondita discussione che Majorana compie sul
concetto di stato. \\
\`E infine da evidenziare come nell'analisi di Majorana giochino
un ruolo fondamentale anche le propriet\`a di simmetria del
sistema fisico.
\begin{quote}
 Sotto certe ipotesi che, nei problemi assai semplici che
esamineremo, sono verificate si pu\`o dire che ogni stato quantico
possiede tutti i caratteri di simmetria dei vincoli del sistema.
\end{quote}
L'attinenza con la formulazione in termini di integrali sui
cammini pu\`o riferirsi a quanto segue. Nel discutere un dato
sistema atomico Majorana sottolinea come da uno stato quantico $S$
del sistema si possa passare ad un altro stato $S'$ mediante
un'operazione di simmetria.
\begin{quote}
 A differenza per\`o di ci\`o che accade nella meccanica
classica per le \ul{singole soluzioni} delle equazioni dinamiche,
non \`e pi\`u vero in generale che $S'$ sia diverso da $S$.
Possiamo rendercene conto facilmente associando a $S'$, come si
\`e visto, un complesso di soluzioni delle equazioni classiche,
poich\`e basta allora che $S$ comprenda con ogni soluzione anche
l'altra che si deduce dalla prima in base a una propriet\`a di
simmetria nei moti del sistema perch\`e $S'$ risulti costituito
identicamente a $S$.
\end{quote}
Tale passaggio \`e peculiarmente intrigante se si osserva che la
questione della ridondanza nella misura d'integrazione per teorie
che esibiscono simmetrie di gauge, e che porta ad espressioni
infinite per l'ampiezza di transizione \cite{ridondanza}, fu
sollevata (e risolta) solo molto tempo dopo il lavoro iniziale di
Feynman. \\
In conclusione, sebbene sia indubbio che nel testo di Majorana non
vi sia alcuna traccia indicativa del formalismo sottostante
all'approccio di Feynman alla meccanica quantistica in termini di
integrali sui cammini (come \`e invece presente nell'articolo di
Dirac del 1933 che forse Majorana conosceva), \`e comunque molto
interessante notare che in esso sono presenti i caratteri fisici
salienti del nuovo modo di interpretare la teoria dei quanti. E
tale circostanza \`e particolarmente significativa se si ha
presente il fatto che, nell'esposizione storica nota,
l'interpretazione del formalismo \`e venuto soltanto dopo lo
sviluppo di quest'ultimo. \\
Nel manoscritto di Majorana, poi, sono presenti anche alcune
interessanti applicazioni a sistemi atomici e molecolari, in cui
risultai noti sono ricavati o reinterpretati secondo il nuovo
punto di vista. Di questi, per\`o, si lascer\`a la scoperta la
lettore nelle pagine che seguono, dove verr\`a riportato
interamente il testo qui discusso.

\section{Il testo della lezione}

\noindent Il manoscritto di Majorana, come pu\`o vedersi in
\cite{Bibliopolis}, presenta al suo inizio un indice numerato
degli argomenti da svolgere, solo parzialmente rispettato nel
testo che segue. Per comodit\`a di lettura si \`e qui preferito
suddividere l'intero scritto in paragrafi secondo l'indice iniziale. \\

%${}$ \\

\setcounter{equation}{0}
\renewcommand{\theequation}{\Roman{equation}}

\renewcommand{\thesubsection}{\arabic{subsection}}

\subsection{Sul significato di stato quantico}

${}$

\noindent L'energia interna di un sistema chiuso (atomo, molecola,
etc.) \`e suscettibile secondo la meccanica quantistica di una
successione discreta di valori $E_0,E_1,E_2,\dots$ che
costituiscono i cosiddetti ``autovalori'' dell'energia. A ogni
determinazione dell'energia corrisponde uno ``stato quantico'',
cio\`e uno stato in cui il sistema pu\`o permanere indefinitamente
in assenza di perturbazioni esterne. Fra queste si deve in
generale considerare l'accoppiamento del sistema con il campo di
radiazione in virt\`u del quale esso pu\`o perdere energia sotto
forma di irraggiamento elettromagnetico passando da un livello
energetico $E_k$ a un livello inferiore $E_i<E_k$. Solo quando
l'energia interna ha il valore minimo $E_0$, essa non pu\`o
diminuire ulteriormente per irradiazione; si dice allora che il
sistema \`e nel suo ``{stato fondamentale}'' da cui non pu\`o
essere rimosso senza l'intervento di influenze
esterne\footnote{Inizialmente Majorana usa il termine tecnico {\it
perturbazioni} in luogo del pi\`u generico {\it influenze}.}
sufficientemente intense, come l'urto con particelle veloci o con
quanti di luce di frequenza elevata.
\\
Che cosa corrisponde nella meccanica classica a uno stato
quantico? A questa domanda si deve anzitutto rispondere se si vuol
giungere a una rappresentazione\footnote{Una stesura precedente
specifica che tale rappresentazione \`e {\it sostanzialmente}
corretta, l'avverbio essendo successivamente cancellato.} corretta
dei risultati a cui \`e pervenuta nel nostro campo la meccanica
quantistica, senza tuttavia entrare a fondo nei complessi metodi
di calcolo da questa adottati. \\
Nella meccanica classica il movimento di un sistema composto di
$N$ punti materiali \`e interamente determinato quando sono note
le coordinate $q_1 \dots q_{3N}$ di tutti punti in funzione del
tempo: %
\be \label{a1} %
q_i = q_i(t)
\ee %
Le (\ref{a1}) risolvono\footnote{Il testo \`e oscuro e
l'interpretazione di tale parola \`e solo probabile.} le equazioni
della dinamica in cui figurano tutte le forze interne ed esterne a
cui il sistema \`e soggetto e possono sempre scegliersi in modo
che per un certo istante tutte le coordinate $q_i(0)$ e le loro
derivate temporali $\dot{q}_i(0)$ abbiano valori arbitrariamente
prefissati. Cos\`i la soluzione generale delle equazioni del
movimento deve contenere $2 \cdot 3 N$ costanti arbitrarie. \\
Per sistemi di proporzioni atomiche la rappresentazione classica
non \`e pi\`u valida e ne sono state proposte successivamente due
modificazioni. La prima, che \`e dovuta a Bohr e Sommerfeld e ha
reso lungamente utili servizi, \`e stata in seguito interamente
abbandonata di fronte al sorgere della nuova meccanica quantistica
che sola ha fornito un formalismo di estrema generalit\`a, e
pienamente confermato dall'esperienza per lo studio dei fenomeni
elementari. Secondo la vecchia teoria di Bohr-Sommerfeld la
meccanica classica \`e ancora valida all'interno dell'atomo cos\`i
che il movimento dell'elettrone ad es. intorno al nucleo di
idrogeno \`e ancora descritto mediante una soluzione (\ref{a1})
delle equazioni della meccanica classica; per\`o se si considerano
movimenti aventi carattere periodico, come appunto la rivoluzione
di un elettrone intorno al nucleo, non tutte le soluzioni delle
equazioni classiche sono realizzate in natura, ma solo
un'infinit\`a discreta di esse soddisfacente alle cosiddette
{condizioni di Sommerfeld}, cio\`e a certe relazioni integrali di
sapore quasi cabalistico. Per esempio in ogni moto periodico in
una dimensione l'integrale esteso a tutto il periodo $\t$:
\[ \int_0^\t \!\! 2 T(t) \, \drm t = n h \]
del doppio dell'energia cinetica deve essere multiplo intero della
{costante di Planck} ($h=6.55 \cdot 10^{-27}$). L'unione della
meccanica classica e di un principio estraneo al suo spirito, come
quello delle orbite quantizzate, appare cos\`i ibrida che non deve
recare meraviglia la rovina completa della teoria avvenuta
nell'ultimo decennio ad onta di varie prove sperimentali in suo
favore che si erano credute
decisive. \\
La nuova meccanica quantistica, dovuta precipuamente a Heisenberg,
\`e sostanzialmente pi\`u prossima alle concezioni classiche che
non la vecchia. Secondo la teoria di Heisenberg uno stato quantico
corrisponde non a una soluzione stranamente privilegiata delle
equazioni classiche ma piuttosto a un complesso di soluzioni
differenti per le condizioni iniziali e anche per l'energia, ci\`o
che si presenta come energia, esattamente definita, dello stato
quantico corrispondendo a una specie di media presa sulle infinite
orbite classiche raccolte in esso. Gli stati quantici si
presentano cos\`i come i minimi complessi statistici di movimenti
classici \ul{poco differenti} accessibili all'osservazione. Questi
complessi statistici minimi non si lasciano ulteriormente dividere
a causa del {principio di indeterminazione}, enunciato dallo
stesso Heisenberg, che vieta la misura esatta contemporanea della
posizione e della velocit\`a
di un particella e cos\`i la fissazione della sua orbita. \\
Un oscillatore armonico di frequenza $\n$ pu\`o oscillare
classicamente con ampiezza e fase arbitrarie, la sua energia
essendo data da
\[ E = 2 \pi^2 m \n^2 A_0^2 \]
se $m$ \`e la sua massa e $A_0$ l'elongazione massima. Secondo la
meccanica quantistica i valori possibili di $E$ sono, come \`e
noto, $\dps E_0 = \frac{1}{2} h \n$, $\dps E = \frac{3}{2} h \n$,
\dots $\dps E_n = \left( n + \frac{1}{2} \right) h \n$ \dots; in
questo caso si pu\`o dire che lo stato fondamentale con energia
$\dps E_0 = \frac{1}{2} h \n$ corrisponde grosso modo a tutte le
oscillazioni classiche con energia compresa fra $0$ e $h \n$, il
primo stato eccitato con energia $\dps E_0 = \frac{3}{2} h \n$
corrisponde alle soluzioni classiche con energia compresa fra $h
\n$ e $2 \cdot h \n$, e cos\`i via. Naturalmente la corrispondenza
fra stati quantici e complessi di soluzioni classiche non \`e che
approssimata poich\`e le equazioni della dinamica quantistica sono
in generale indipendenti dalle corrispondenti equazioni classiche
e indicano una reale modificazione delle leggi meccaniche oltre
che un limite posto alle possibilit\`a di osservazione; ma essa
\`e meglio fondata della rappresentazione degli stati quantici
mediante orbite quantizzate e pu\`o essere utilmente impiegata in
considerazioni di carattere qualitativo.

\subsection{Le propriet\`a di simmetria di un sistema nella meccanica
classica e quantistica}

${}$

\noindent Uno studio particolare meritano i sistemi aventi qualche
carattere di simmetria. Per essi da una soluzione particolare
delle equazioni classiche del movimento $q_i = q_i(t)$ se ne
deducono in generale altre differenti $q_i' = q_i'(t)$ in base a
sole considerazioni di simmetria. Per esempio se il sistema
contiene due o pi\`u elettroni o in genere due o pi\`u particelle
uguali si pu\`o, scambiando le coordinate di due particelle
passare da una soluzione a un'altra che sar\`a in generale
differente. Analogamente se consideriamo un elettrone mobile nel
campo di due nuclei o atomi uguali (nella figura $A$ e $B$) si
pu\`o, riflettendo nel centro $O$ del segmento $AB$, dedurre da
un'orbita possibile $q_i = q_i(t)$ percorsa con determinata legge
oraria intorno ad $A$, un'orbita $q_i' = q_i'(t)$  descritta
dall'elettrone intorno al nucleo o atomo $B$.
\begin{center}
%\begin{figure}[t]
\epsfysize=2cm \epsfxsize=6.5truecm %
\centerline{\epsffile{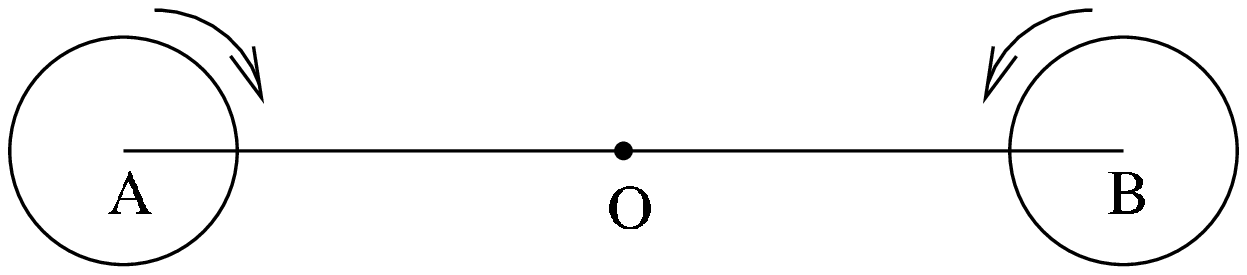}}
%\caption{}
%\label{app1fig}
%\end{figure}
\end{center} \vspace{-0.5truecm}
Le operazioni di scambio di due particelle uguali, riflessioni in
un punto, o altre corrispondenti a caratteri qualsiasi di
simmetria, conservano il loro significato nella meccanica
quantistica. Cos\`i \`e possibile dedurre da uno stato quantico
$S$ un altro stato $S'$, \ul{appartenente allo stesso noto valore
dell'energia}, se nei due esempi citati si scambiano due
particelle uguali e rispettivamente\footnote{Vedi nota
precedente.} si riflette il sistema nel punto $O$. A differenza
per\`o di ci\`o che accade nella meccanica classica per le
\ul{singole soluzioni} delle equazioni dinamiche, non \`e pi\`u
vero in generale che $S'$ sia diverso da $S$. Possiamo rendercene
conto facilmente associando a $S'$, come si \`e visto, un
complesso di soluzioni delle equazioni classiche, poich\`e basta
allora che $S$ comprenda con ogni soluzione anche l'altra che si
deduce dalla prima in base a una propriet\`a di simmetria nei moti
del sistema perch\`e $S'$
risulti costituito identicamente a $S$. \\
In alcuni casi se il sistema presenta caratteri di simmetria
sufficientemente complessi \`e in realt\`a possibile ottenere per
simmetria da uno stato quantico altri stati diversi ma aventi la
stessa energia. Si dice allora che il sistema \`e \ul{degenere},
ha cio\`e parecchi stati di uguale energia, in virt\`u appunto dei
suoi caratteri di simmetria. Lo studio dei sistemi degeneri e
delle condizioni in cui pu\`o aversi degenerazione ci porterebbe
troppo lontano ed \`e del resto difficile a eseguirsi in base a
sole analogie classiche. Lo lasceremo peci\`o interamente da parte
confinando la nostra attenzione su problemi in cui non si presenta
degenerazione. Questa condizione \`e sempre soddisfatta se la
simmetria del sistema meccanico consente solo una trasformazione
cos\`i semplice che il suo quadrato, cio\`e la sua ripetizione due
volte consecutivamente, si riduce a una trasformazione identica.
Per esempio, riflettendo due volte di seguito un sistema di punti
materiali rispetto a un piano, a una retta o a un punto ritroviamo
necessariamente la stessa disposizione iniziale; analogamente
scambiando due volte due particelle uguali il sistema rimane
inalterato. In tutti questi casi si presentano solo stati quantici
semplici, cio\`e a ogni valore possibile dell'energia appartiene
un solo stato quantico. Segue che tutti gli stati quantici di un
sistema contenente due particelle uguali sono simmetrici nelle due
particelle, restando inalterati per il loro scambio. E cos\`i gli
stati di un elettrone circolante intorno a due nuclei uguali $A$ e
$B$ sono simmetrici rispetto al punto medio $O$ di $AB$, o restano
inalterati per riflessione in $O$, e analogamente in altri casi
simili. Sotto certe ipotesi che, nei problemi assai semplici che
esamineremo, sono verificate si pu\`o dire che ogni stato quantico
possiede tutti i caratteri di simmetria dei vincoli del sistema.

\subsection{Forze di risonanza fra stati non simmetrizabili per
perturbazione piccola e conseguenze spettroscopiche. Teoria della
valenza omeopolare secondo il metodo degli elettroni leganti.
Propriet\`a degli stati simmetrizzati che non si ottengono per
perturbazione debole da stati non simmetrizzati}

\noindent \footnote{Nel manoscritto originale ci\`o che viene
raggruppato sotto un unico paragrafo corrisponde grossolanamente a
quanto nell'indice viene indicato con tre diversi paragrafi: \\
3) Forze di risonanza fra stati non simmetrizabili per
perturbazione piccola. Caratteri di simmetria non combinabili. \\
4) Conseguenze spettroscopiche in atomi con due elettroni.
Risonanza fra buche uguali di potenziali e teoria della valenza
omeopolare secondo il metodo degli elettroni leganti. \\
5) Propriet\`a degli stati simmetrizzati che non si ottengono per
perturbazione debole da stati non simmetrizzati. Bande alternate,
idrogeno \dots.}

\noindent Consideriamo ora un elettrone mobile nel campo di due
nuclei di idrogeno o protoni. Il sistema dei due protoni e
dell'elettrone ha una carica netta risultante $+e$, e costituisce
la pi\`u semplice molecola realizzabile, cio\`e la molecola di
idrogeno ionizzata positivamente. In tale sistema sono mobili
cos\`i i protoni come l'elettrone, ma a causa della grande
differenza di massa fra i primi e il secondo (rapporto delle masse
1840:1) la velocit\`a media dei protoni \`e molto inferiore a
quella dell'elettrone, e il movimento di quest'ultimo si pu\`o
studiare con grande approssimazione supponendo i protoni fermi a
una certa distanza reciproca. Questa distanza \`e determinata da
ragioni di stabilit\`a in modo che sia minima l'energia totale
della molecola, che risulta in prima approssimazione dalla somma
dell'energia potenziale mutua dei due protoni e dell'energia
dell'elettrone mobile nel campo dei primi due, ed \`e diversa per
i diversi stati quantici dell'elettrone. \\
L'energia potenziale mutua dei protoni \`e data da $\dps
\frac{e^2}{r}$ se $r$ \`e la loro distanza, mentre l'energia di
legame dell'elettrone nel suo stato fondamentale \`e una funzione
negativa $E(r)$ di $r$ che non ha un'espressione analitica
semplice ma pu\`o essere calcolata dalla meccanica quantistica con
approssimazione grande quanto si desideri. La distanza $r_0$ di
equilibrio \`e cos\`i determinata dalla condizione che sia minima
l'energia totale
\[ W(r_0) = \frac{e^2}{r_0} + E(r_0) \]
La curva $W(r)$ ha un andamento del tipo indicato in figura, se si
assume come zero dell'energia quella corrispondente alla molecola
dissociata in un atomo neutro di idrogeno e un atomo ionizzato a
distanza infinita. La distanza di equilibrio \`e stata
\begin{center}
%\begin{figure}[t]
\epsfysize=3.5cm \epsfxsize=6cm %
\centerline{\epsffile{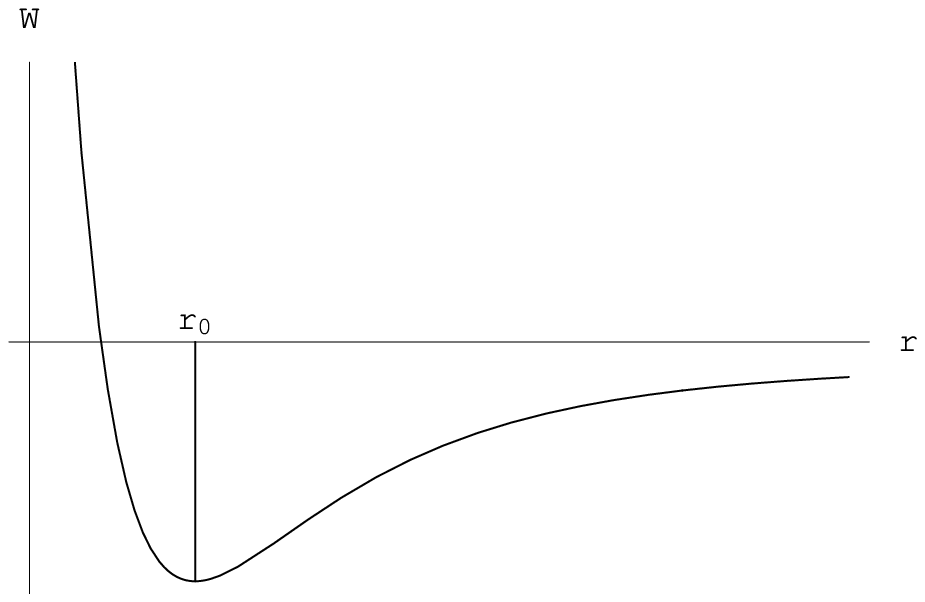}}
%\caption{}
%\label{app2fig}
%\end{figure}
\end{center} \vspace{-0.5truecm}
calcolata teoricamente da Burrau\footnote{Majorana si riferisce
qui all'articolo di \O. Burrau, {\it Berechnung des Energiewertes
des Wasserstoffmolekel-Ions (H2+) im Normalzustand}, Kgl. Danske
Videnskab. Selskab, Mat. Fys. Medd. {\bf 7} (1927) 14.} che ha
trovato $r_0 = 1.05 \cdot 10 ^{-8}$ cm e per l'energia
corrispondente $W(r_0) = - 2.75$ volt-elettrone. Questi risultati
sono stati entrambi pienamente confermati dall'osservazione a cui
sono indirettamente accessibili attraverso lo studio dello spettro
emesso dalla molecola neutra o
ionizzata. \\
Qual'\`e l'origine della forza $\dps F = + \frac{\drm W}{\drm r}$
che tende ad avvicinare i due nuclei di idrogeno quando essi si
trovano a una distanza maggiore di $r_0$? La risposta che la
meccanica quantistica d\`a a tale questione \`e sorprendente
poich\`e sembra indicare che accanto a certe {forze di
polarizzazione} prevedibili in base alla meccanica classica hanno
una parte preponderante forze di un tipo assolutamente nuovo, le
cosiddette {forze di risonanza}. \\
Supponiamo la distanza $r$ grande rispetto al raggio dell'atomo
neutro di idrogeno ($\sim 0.5 \cdot 10^{-8} cm$). L'elettrone si
trova allora sotto l'azione dell'uno o dell'altro dei due protoni
intorno a ognuno dei quali pu\`o descrivere classicamente  orbite
chiuse. Il sistema formato dall'elettrone e dal nucleo intorno a
cui si muove costituisce un atomo neutro di idrogeno, cos\`i che
la nostra molecola risulta essenzialmente composta da un atomo
neutro e da un protone a una certa distanza dal primo. L'atomo
neutro di idrogeno nel suo stato fondamentale ha una distribuzione
di carica a simmetria sferica, ci\`o che classicamente significa
che tutte le orientazioni dell'orbita elettronica sono egualmente
possibili, la densit\`a di carica negativa decrescendo
esponenzialmente con la distanza in modo che il raggio dell'atomo
si presenta praticamente come finito; segue che nessun campo
elettrico \`e generato all'esterno da un atomo neutro di idrogeno
e cos\`i nessuna azione pu\`o essere esercitata su un protone
posto a una distanza $r$ grande rispetto alle dimensioni
dell'atomo. In realt\`a per\`o l'atomo neutro si polarizza sotto
l'azione del protone esterno e acquista un momento elettrico nella
direzione protone - atomo neutro, e dall'interazione di questo
momento elettrico con il campo non uniforme generato dal protone
sorge una forza attrattiva che tende ad associare atomo e ione in
un sistema molecolare. \\
Le {forze di polarizzazione} facilmente prevedibili con
considerazioni classiche posso dare origine da sole a composti
molecolari, le cosiddette molecole di polarizzazione, che sono
per\`o caratterizzate da un'estrema labilit\`a. Complessi
molto\footnote{Majorana usa inizialmente il rafforzativo {\it
assai pi\`u} invece di {\it molto}, successivamente cancellato.}
stabili possono aversi solo se altre forze si aggiungono a quelle
di polarizzazione. Nelle {molecole polari}, composte di due ioni
di segno differente, queste forze sono date essenzialmente
dall'attrazione elettrostatica fra gli ioni; ad es. la molecola
$HCl$ \`e tenuta insieme essenzialmente dall'attrazione mutua fra
lo ione positivo $H^+$ e lo ione negativo $Cl^-$. Ma in molecole
composte di due atomi neutri, o di un atomo neutro e un atomo
ionizzato, come nel caso dello ione molecolare $H_2^+$,
l'{affinit\`a chimica} \`e essenzialmente condizionata dal
fenomeno di risonanza nel significato, senza analogia nella
meccanica classica, che questa parola ha assunto nella nuova
meccanica. \\
Quando si studia dal punto di vista della meccanica quantistica il
movimento dell'elettrone nel campo dei due protoni supposti fissi
a una distanza mutua $r$ molto grande si possono in prima
approssimazione determinarne i livelli energetici supponendo che
l'elettrone debba muoversi intorno al protone $A$ (o $B$) e
prescindendo dall'esistenza dell'altro protone $B$ (o $A$) che
esercita un'azione perturbativa debole a causa della sua distanza.
Si ottiene cos\`i per il pi\`u basso autovalore dell'energia $E_0$
uno stato $S$ che corrisponde alla formazione di un atomo neutro
nello stato fondamentale fra l'elettrone e il nucleo $A$ e uno
stato $S'$ che corrisponde a un atomo neutro formato fra
l'elettrone e il nucleo $B$. Se ora si tiene conto della
perturbazione che in ognuno dei due casi \`e esercitata sull'atomo
neutro dallo ione positivo noi troveremo ancora, finch\`e la
perturbazione \`e piccola, in luogo di due autovalori uguali $E_0$
due autovalori $E_1$ e $E_2$ poco differenti perch\`e entrambi
prossimi a $E_0$; ma gli stati quantici ad essi corrispondenti, e
siano essi $T_1$ e $T_2$, non sono prossimi separatamente a $S$ e
$S'$, perch\`e essendo il campo di potenziale in cui l'elettrone
si muove simmetrico rispetto al punto medio di $AB$, la stessa
simmetria devono presentare, per quanto si \`e detto prima, gli
stati effettivi $T_1$ e $T_2$ dell'elettrone, mentre non la
presentano separatamente $S$ e $S'$.\\
Secondo la rappresentazione modellistica degli stati quantici che
abbiamo descritto pi\`u sopra, $S$ \`e composto da un complesso di
orbite elettroniche circolanti intorno ad $A$, $S'$ analogamente
da un complesso di orbite intorno a $B$, mentre gli stati quantici
veri del sistema $T_1$ e $T_2$ abbracciano ognuno, in prima
approssimazione per $r$ molto grande, met\`a delle orbite
costituenti $S$ e met\`a di quelle che costituiscono $S'$. Il
calcolo prova che per distanze nucleari sufficientemente grandi il
valor medio degli autovalori perturbati $E_1$ e $E_2$ coincide
sensibilmente con l'unico valore imperturbato $E_0$, mentre la
loro differenza \`e apprezzabile e ha importanza decisiva in
questo e in infiniti altri casi analoghi per lo studio delle
reazioni chimiche. Si pu\`o dunque supporre $E_1 < E_0$ ma $E_2 >
E_0$ e sar\`a allora $T_1$ lo stato fondamentale dell'elettrone
mentre $T_2$ apparir\`a come stato eccitato con energia
leggermente superiore. \\
L'elettrone nello stato $T_1$, come pure nello stato $T_2$, passa
met\`a del suo tempo intorno al nucleo $A$ e l'altra met\`a
intorno al nucleo $B$. Si pu\`o anche stimare la frequenza media
del passaggio periodico dell'elettrone da $A$ a $B$ e viceversa, e
cos\`i dallo scambio dello stato, neutro o ionizzato, dei due
atomi, e si trova
\[ \n = \frac{E_2 - E_1}{h} \]
essendo $h$ la costante di Planck. Per grandi valori di $r$, $E_2
- E_1$ decresce secondo una curva di tipo esponenziale e cos\`i la
frequenza di scambio tende rapidamente a zero, e questo significa
che l'elettrone posto inizialmente intorno ad $A$ vi rimane per un
tempo sempre pi\`u lungo, ci\`o che \`e ben comprensibile da un
punto di vista classico. \\
Se l'elettrone si trova nello stato $T_1$, cio\`e nel suo stato
fondamentale, la sua energia ($E_1$) \`e inferiore a quella
($E_2$) che esso avrebbe senza l'accennato effetto di scambio fra
l'ufficio dei nuclei $A$ e $B$. Questa circostanza\footnote{Vedi
nota 5.} d\`a origine a un nuovo tipo di forze attrattive fra i
nuclei che si aggiungono alle forze di polarizzazione considerate
prima e sono precisamente la causa determinante dell'unione
molecolare. \\
Le forze di risonanza non hanno come abbiamo detto analogia nella
meccanica classica. In realt\`a fin dove pu\`o condurci l'analogia
che ci ha permesso di associare a uno stato quantico un complesso
statistico di movimenti classici, i due stati $T_1$ e $T_2$ in cui
pure le forze di risonanza hanno segno opposto risultano
identicamente costituiti, ognuno dalla met\`a di entrambi
gli\footnote{Vedi nota 5.} originari stati imperturbati $S$ e
$S'$. Ma questo \`e vero soltanto in una certa approssimazione,
cio\`e nell'approssimazione stessa in cui sono trascurabili le
forze di risonanza. Per un calcolo esatto che tenga conto delle
forze di risonanza si deve usare necessariamente la meccanica
quantistica e si trova allora una differenza \ul{qualitativa}
nella struttura dei due stati quantici che si manifesta
soprattutto nella regione intermedia fra $A$ e $B$ attraverso la
quale con un meccanismo non descrivibile secondo la meccanica
classica ha luogo il passaggio periodico dell'elettrone da un
atomo all'altro. Questa differenza qualitativa ha un carattere
assolutamente formale e possiamo enunciarla solo attraverso
l'introduzione della funzione d'onda $\psi(x,y,z)$ che, come \`e
noto, \`e associata nella meccanica quantistica ad ogni stato
possibile dell'elettrone. Il modulo del quadrato di $\psi$, che
pu\`o anche essere una grandezza complessa, d\`a la probabilit\`a
che l'elettrone si trovi nell'unit\`a di volume intorno a un punto
generico $x,y,z$. La funzione d'onda $\psi$ deve allora soddisfare
a un'equazione differenziale lineare e cos\`i \`e sempre lecito
moltiplicare $\psi$ in tutti i punti per uno stesso numero reale o
complesso di modulo $1$, quest'ultima limitazione essendo
richiesta dalla condizione di normalizzazione
\[ \int \! | \psi^2 | \, \drm x \, \drm y \, \drm z = 1 \]
necessaria per l'accennata interpretazione fisica di $|\psi^2|$.
La moltiplicazione di $\psi$ per una costante di modulo $1$ lascia
inalterata la distribuzione spaziale della carica elettronica e
non ha in generale alcun significato fisico. Noi definiremo ora
formalmente la riflessione di uno stato quantico nel punto $O$
medio dei due nuclei $A$ e $B$ direttamente sulla funzione d'onda
$\psi$, ponendo
\[ \psi(x,y,z) = \psi'(-x,-y,-z) \]
in un sistema di coordinate aventi l'origine in $O$. Se $\psi$
deve rappresentare uno stato quantico simmetrico e quindi
invariante per riflessione in $O$, la funzione d'onda riflessa
$\psi'$ deve avere lo stesso significato fisico di $\psi$ e cos\`i
differire da $\psi$, per quanto si \`e detto, per un fattore
costante, reale o complesso di modulo $1$. Questo fattore costante
deve inoltre\footnote{Vedi nota 5.} valere $\pm 1$, poich\`e il
suo quadrato deve dare l'unit\`a come prova il fatto riflettendo
ancora $\psi'$ nel punto $O$ si ottiene di nuovo la funzione
d'onda iniziale $\psi$. \\
Per tutti gli stati del nostro sistema si dovr\`a dunque avere:
\[ \psi (x,y,z) = \pm \psi (-x,-y,-z) \]
dovendosi scegliere il segno $+$ per una parte di essi e il segno
$-$ per gli altri. La differenza formale fra gli stati $T_1$ e
$T_2$ considerati pi\`u sopra consiste precisamente in
questo\footnote{Nell'originale compaiono qui di seguito altre tre
parole aggiunte successivamente, la cui interpretazione \`e
oscura.} che nell'equazione precedente per $T_1$ \`e valido il
segno superiore ma per $T_2$ quello inferiore. La simmetria
rispetto a un punto, e cos\`i in genere ogni propriet\`a di
simmetria, determina una distinzione formale degli stati del
sistema in due o pi\`u categorie, una propriet\`a importante di
questa distinzione essendo che non possono provocarsi transizioni
fra stati appartenenti a categorie diverse mediante perturbazioni
esterne che rispettino le simmetrie esistenti nei vincoli del
sistema. Cos\`i in sistemi contenenti due elettroni si hanno due
categorie di stati non combinabili distinti secondo che la
funzione d'onda, che ora dipende dalle coordinate di entrambi gli
elettroni, resta inalterata o cambia segno scambiando le due
particelle uguali. Nel caso speciale dell'atomo di elio questo
d\`a luogo alla ben nota apparenza spettroscopica di due elementi
distinti: parelio e {ortoelio}. \\
La teoria dell'affinit\`a chimica fra atomo neutro e atomo
ionizzato di idrogeno che abbiamo considerato fin qui pu\`o essere
estesa per lo studio della molecola neutra di idrogeno e pi\`u in
generale di tutte le molecole risultanti da due atomi neutri
uguali. Per la molecola neutra di idrogeno dovremo considerare in
luogo di un solo elettrone mobile intorno a due protoni fissi,
\ul{due} elettroni mobili nello stesso campo prescindendo in una
prima approssimazione dalla loro mutua repulsione. La stabilit\`a
della molecola \`e allora comprensibile assumendo che ognuno dei
due elettroni cada nello stato $T_1$ a cui corrispondono forze di
risonanza attrattive. Si pu\`o dire con F. Hund che la molecola di
idrogeno \`e tenuta insieme da due elettroni ``leganti''. In
realt\`a l'interazione degli elettroni \`e cos\`i forte da
lasciare solo una giustifica\footnote{Vedi nota 5.} qualitativa
alla teoria schematica di Hund, ma nulla vieta in principio di
prevedere con esattezza tutte le propriet\`a della molecola di
idrogeno ... \footnote{Il testo qui omesso \`e completamente
oscuro.} risolvendo con sufficiente precisione le equazioni poste
dalla meccanica quantistica. Cos\`i si \`e potuto, con metodi
matematici appropriati, determinare effettivamente\footnote{Vedi
nota 5.} in base a pure considerazioni teoriche l'affinit\`a
chimica fra due atomi neutri di idrogeno e il valore teorico si
accorda con quello sperimentale, entro i limiti di approssimazione
in cui, per ragioni pratiche, si \`e voluto condurre il calcolo. \\
Per molecole diverse da quella di idrogeno\footnote{Nella prima
stesura dell'originale, poi cancellata, compare: {\it Per molecole
composte da atomi pi\`u pesanti \dots}.} la teoria dell'affinit\`a
chimica \`e notevolmente pi\`u complicata sia per il maggior
numero di elettroni che occorre considerare, sia a causa del
principio di Pauli che vieta la contemporanea presenza di pi\`u di
due elettroni nello stesso stato; ma le diverse teorie
dell'affinit\`a chimica che sono state proposte negli ultimi anni
e di cui ognuna ha un campo pi\`u o meno largo di validit\`a,
consistono in sostanza nella ricerca di metodi di calcolo
approssimati per la risoluzione di un problema matematico che \`e
in s\`e esattamente determinato, non gi\`a
nell'enunciazione di nuovi principi fisici. \\
Cos\`i \`e possibile ricondurre la teoria delle saturazioni delle
valenze ai pi\`u generali principi della fisica. La meccanica
quantistica apre\footnote{Nel manoscritto originale compare qui un
avverbio non identificato che sostituisce il {\it finalmente} di
una stesura preliminare.} la strada per l'unificazione logica di
tutte le scienze che hanno per oggetto comune il mondo inorganico.

\section*{Ringraziamenti}

\noindent Sono molto grato al Dr. Alberto De Gregorio per le
discussioni e i preziosi suggerimenti offerti, nonch\`e per una
attenta lettura di una versione preliminare del presente articolo.
Ringrazio inoltre il Prof. Erasmo Recami per la fattiva
collaborazione e per numerose discussioni.

\vspace{1truecm}

% ----------------------------------------------------------------

\end{document}